\documentclass[aps,superscriptaddress]{revtex4}

\usepackage{amsmath,amsfonts,amssymb}
\usepackage{graphicx}

\begin{document}

\title{Vortex in axion condensate as a dark matter halo}

\author{Jakub Mielczarek\footnote{jakub.mielczarek@uj.edu.pl}}
\affiliation{Astronomical Observatory, Jagiellonian University, 30-244
Krak\'ow, ul. Orla 171, Poland}

\author{Tomasz Stachowiak\footnote{t.stachowiak@gmail.com}}
\affiliation{Astronomical Observatory, Jagiellonian University, 30-244
Krak\'ow, ul. Orla 171, Poland}

\author{Marek Szyd{\l}owski\footnote{uoszydlo@cyf-kr.edu.pl}}

\affiliation{Astronomical Observatory, Jagiellonian University, 30-244
Krak\'ow, ul. Orla 171, Poland}
\affiliation{Marc Kac Complex Systems Research Centre, Jagiellonian University,
ul. Reymonta 4, 30-059 Krak{\'o}w, Poland}

\begin{abstract}
We study the possibility of the vortices formation in axion condensate on the
galactic scale. Such vortices can occur as a result of global rotation of the 
early universe. We study analytical models of vortices and calculate exemplary 
galaxy rotation curves. Depending on the setup it is possible to obtain a 
variety of shapes which give a good qualitative agreement with observational 
results. However, as we show, the extremely low velocity dispersions of the 
axions velocity are required to form the single vortex on the galactic scales. 
We find that the required velocity dispersion is of the order of 
$\sigma \approx 10^{-12} \frac{m}{s}$. This is much smaller that predicted 
within the present understanding of the axion physics. The vortices in axion 
condensate can be however formed on the much smaller scales and give seeds to 
the galaxy formation and to their angular momenta. On the other hand, the 
vortices can be formed on the galactic scales, but only if the mass of the 
axion-like particles is of the order of $10^{-30}$eV.  In this case, the 
particle de Broglie wavelength is comparable with the galactic diameter. 
This condition must be fulfilled in order to keep the coherence of the quantum 
condensate on the galactic scales.
\end{abstract}

\maketitle

\section{Introduction}

The problem of dark matter seems to be a challenge for theoretical and 
observational investigations. Recent astronomical observations, like the 
measurements of cosmic microwave background (CMB) radiation by 
WMAP \cite{vortex:WMAP}, or measurements of distant supernova type 
Ia (SNIa) \cite{vortex:Riess,vortex:Perlmutter}, indicate that the Universe, 
apart from matter, is fulfilled with a phenomenological fluid with negative 
pressure (called dark energy) which could be responsible for the current 
acceleration of the universe. While the nature of this energy is still unknown 
(the cosmological constant $\Lambda$ is the most serious candidate), the 
combination of results from CMB measurements, SNIa data and extragalactic 
observations indicate that as much as 2/3 of total energy density of the 
Universe is in the form of the mysterious dark energy. While it dominates 
the dynamics of the Universe on the large scale, dark matter (matter 
whose existence has been inferred only through its gravity) clearly 
influences the galactic scale dynamics. Its abundance is given in terms 
of the density parameter $\Omega_{\mathrm{DM}}=\rho_{\textrm{DM}}/\rho_{\text{c}}$, 
where $\rho_{\text{c}}$ is the critical energy density, $\rho_{\text{c}}=\frac{3H^2}{8\pi G}$. 
For the ``concordance'' flat $\Lambda$CDM model we have total matter 
$\Omega_{\mathrm{m}}\simeq0.3$; in turn from emission and absorption of 
photons visible matter is roughly $\Omega_{\textrm{vis}} \simeq 0.04$ and 
it gives us that dark matter amounts to $\Omega_{\textrm{DM}} \simeq 0.26 $.

The rotation curves of spiral galaxies give us the strongest evidence for dark
matter. This dark matter in its $80\%$ is in some form cold and nonbaryonic.
Therefore, we can identify only $4\%$  of the total matter content in the present 
universe. If we do not postulate the existence of dark matter it is not possible 
to explain why at large distance $r$ from the centre of a given galaxy,
we would find circular velocity $v^2_c \simeq G M_{\mathrm{vis}}/r$,
since visible matter is concentrated around its centre. For the observational results of
rotation curves see Ref.~\cite{vortex:Salucci}. However, observations show that
$v_c$ is independent of $r$ at large distances ($v_c \sim 200$km s$^{-1}$ is its
typical value). From numerical simulations of the halo formation we obtain density
profiles for both small and large values $\rho_{\mathrm{halo}}
\propto r^{-\alpha}$ with $\alpha \in (1; 1.5)$ for small and
$\alpha = 3$ for large distances \cite{vortex:More,vortex:Navarro}.
The flat part of the rotation curves corresponds to $\rho_{\mathrm{halo}}
\propto r^{-2}$.  Such a behavior can be explained e.g. within the isothermal 
sphere model of dark matter halo. In this model, the constant value of 
the rotation velocity $v_c$ can be related with the velocity dispersion $\sigma$ of the 
dark matter particles by $v_c=\sqrt{2}\sigma$.

There are two main candidates for cold dark matter, namely the axion and
neutralino \cite{vortex:Lazarides,vortex:Gondolo,Visinelli:2009zm,Hwang:2009js}. In this paper 
we concentrate on the axion.  Axions were originally proposed to solve the strong CP 
problem in the quantum  chromodynamics (QCD). If axions have low mass, thus preventing other decay 
modes, axion theories predict that the universe would be filled with cold Bose-Einstein condensate
of primordial axions \cite{Sikivie:2009qn,Mielke:2009zza}. The axions in this condensate are 
always nonrelativistic and if the mass is about $10^{-3}$ eV it would plausibly 
explain the dark matter problem. There are prospects for direct experimental 
detection of axion. The Axion Dark Matter Experiment (ADMX) \cite{vortex:Duffy} 
searches for weakly interacting axions in the dark matter halo of our galaxy. 
Unfortunately, studies of axion dark matter are not sufficiently sensitive to 
probe the mass regions where axions would be expected.

We investigate the possibility of explanation of flat velocity curves of spiral
galaxies in terms of axion condensate which can be present in the Universe
since the Peccei-Quinn phase transition.

In our investigation we use the Gross-Pitaevski equation in an expanding FRW universe
\begin{equation}
i\hslash \left( \frac{\partial }{\partial t}+\frac{3}{2}\frac{\dot{a}(t)}{a(t)} \right) \phi({\bf r},t)=
 \left(-\frac{\hslash^2}{2m}\frac{1}{a^2(t)}\nabla^2+U({\bf r})+g^2|\phi({\bf r},t)|^2\right)\phi({\bf r},t),
\label{G-P}
\end{equation}
where $U({\bf r})$ is the external potential, $g^2$ is the coupling constant
between axions and $a(t)$ is the scale factor.
Here we describe the condensate by one particle wave function $\phi({\bf r},t)$. 
The circulation in condensate is expressed by \cite{vortex:Ghosh}
\begin{equation}
\Gamma=\oint_{C}{\bf v}\cdot d{\bf r}=\frac{\hslash}{m}2\pi l,
\end{equation}
where $l$ is an integer called topological charge. $C$ denotes any contour
around a vortex. When there is no vortex, $l=0$ and the circulation vanishes. When the
condensate is inside a rotating environment the vortex is formed. In the
interacting condensate vortices with $l>1$ are unstable and decay to the
vortices with $l=1$. So in a realistic situation we have a net of elementary
vortices ($l=1$) which are stable. Such nets of vortices are observed in
laboratories \cite{vortex:Chevy}.

We study the possibility that a galactic halo is just such a vortex. We suppose 
that such a mechanism can be realised in the early universe. Namely, in the 
presence of global rotation, the proposed mechanism yields a huge amount of 
small vortices whose topological charges equal $1$. In this paper we consider 
a singular vortex in the axion condensate. 

The question whether rotation is an attribute of the Universe as a whole has 
been investigated since classical works of Lanczos \cite{vortex:Lanczos}, 
Gamov \cite{vortex:Gamov}, Godel \cite{vortex:Godel}, Hawking \cite{vortex:Hawking} 
and recently by Chapline and Mazur \cite{vortex:Mazur}. When compared with the 
CMB anisotropies, the effects of rotation should not be big today \cite{vortex:Silk}. 
Barrow et al. \cite{vortex:Barrow} showed that the cosmic vorticity depends strongly
on the cosmological model and for a flat universe the bound on vorticity relative to 
the Hubble parameter at present is $\omega/H=2\times 10^{-5}$.  

\section{Free axion condensate}

The Bose-Einstein occurs when the ground state becomes  occupied by the macroscopic 
number of particles. Since each of these particles is in the same quantum state $\phi$, the density
of the condensate composed of the $N$ particles of mass $m$ can be expressed as follows 
$\rho = m N |\phi|^2$.  Therefore, the square modulus of the wave function can be observed just 
as the density distribution of the particles in the condensate. The condition for condensation to occur 
is that de Broglie wavelength of the particles must be comparable with the inter-particles distances. 
The individual wavefunctions of the particles will overlap then. This requirement can be translated 
into the critical temperature $T_c$ below which the condensation starts. For $T<T_c$  phase 
space density occupation number becomes 
significant. It is also important to note that condensation does not occur for the photons (and other 
massless particles) because number of photons is not conserved. Therefore during a cooling, the 
number photons can be e.g. absorbed by the environment and condensation does not take a place.

The axion field can be regarded as massless at the temperatures $T \geq 1$ GeV. In turn, at the lower 
temperatures, $T \leq 1$ GeV, the axion field becomes massive due to the nonperturbative QCD effects.
The oscillations of this field are interpreted as the massive axion particles. The axion mass can be 
expressed as follows \cite{vortex:Sikivie}   
\begin{equation}
m_a \simeq 6 \mu \text{eV}  \left(\frac{10^{12}\ \text{GeV}}{f_a} \right),
\end{equation}
where  $f_a$ is the energy of $U_{PQ}(1)$ symmetry breaking and creation of axions. 
This the $f_a$  is a free parameter, however its value is empirically constrained to 
$10^{9}\ \text{GeV} \lesssim f_a \lesssim 10^{12}$ GeV
This corresponds to the axion mass window 
$6\cdot 10^{-6} \text{eV} \lesssim m_a \lesssim  6 \cdot 10^{-3}$eV. In the subsequent part
of the paper we restrict to the case $f_a = 10^{12}$ GeV what leads to $m_a \approx 10^{-5}$eV.

Because axions are massless
ate the energies  $T \geq 1$ GeV, the condensation cannot take place until the Universe cools down to 
$T \sim 1$ GeV. As it was shown in Ref.~\cite{Sikivie:2009qn}, the axion temperature is then much 
smaller than $T_c$ what allows for the Bose-Einstein condensation to occur.  Namely the temperature is
of the order of the Hubble expansion rate $T \sim 10^{-3}$ eV while  $T_c \sim 300$ GeV. Since the axions 
are very weakly coupled, they can be consider as a gas of free particles. Therefore in the first approximation
the axion condensate can be described by the Gross-Pitaevski equation (\ref{G-P}) with $g=0$ and $U({\bf r})=0$.
However, in order to trace the thermalization of the Bose-Einstein condensate, the weak gravitational 
coupling between axions must be also taken into account. In this section, we neglect this coupling and 
consider the free axion condensate formed at the QCD epoch for $T \sim 1$ GeV. Namely, when the axion
acquired a mass and the conditions for the Bose-Einstein condensation were satisfied.

In this case, the free Gross-Pitaevski equation (\ref{G-P}) takes the form
\begin{equation}
i\hslash \left( \frac{\partial }{\partial t}+\frac{3}{2}\frac{\dot{a}(t)}{a(t)} \right) \phi({\bf r},t) =
-\frac{\hslash^2}{2m}\frac{1}{a^2(t)}\left[  \frac{\partial^2}{\partial r^2}+\frac{2}{r}\frac{\partial}{\partial r}-
\frac{l(l+1)}{r^2}  \right]\phi({\bf r},t). 
\end{equation}
The general solution of this equation is in the form 
\begin{equation}
\phi({\bf r},t)= \frac{C}{a^{3/2}} \exp{\left\{-\frac{i}{\hslash} \int \frac{\mu}{a^2(t)} dt \right\}}
j_l(\sqrt{\lambda}r)  Y_l^n(\theta,\varphi),
\label{sol1}
\end{equation}
where $C$ is the normalization constant and $\lambda=2m\mu/\hslash^2$. Energy of the particle
is defined as $E=\mu/a^2$.  Here $j_l(x)$ is the spherical Bessel function related to the ordinary one with
\begin{equation}
j_l(x)=\sqrt{\frac{\pi}{2x}}J_{l+\frac{1}{2}}(x).
\end{equation}
and $Y_l^n$ are the standard spherical harmonics. Based on (\ref{sol1}), the energy density of the 
axion condensate, composed of $N$ axions can be written as follows
\begin{equation}
\rho({\bf r},t)= m N |\phi({\bf r},t)|^2= \frac{m N |C|^2}{a^{3}} j^2_l(\sqrt{\lambda}r)  |Y_l^n(\theta,\varphi)|^2, 
\end{equation}
where $m$ is axion mass. As we see, during the cosmological evolution, the axion condensate is 
diluted as the ordinary dust matter, namely $\rho \sim 1/a^3$. 

Solution (\ref{sol1}) is non-normalisable, therefore the constant $C$ cannot be directly 
evaluated. To obtain a normalizable solution we must
introduce proper boundary conditions, for example $ \phi(r \geq \bar{r})=0$ for some
$\bar{r}$. But it is not natural in our case, as our Universe is filled 
homogeneously by the axion condensate. Free axions should thus be described
approximately by a non-normalisable waves with defined energies. Accordingly,
our solution describes an axion condensate with defined nonquantized energy.

Our main idea is that small vortices created in the early universe can grow
during expansion of the universe.  At the early stage, the vortex is a quantum 
object but during expansion it can becomes a classical object. The expansion 
of the free condensate follows directly the expansion of the universe. This might
not be a case for the self-interacting condensate. Then, the interaction resists the
expansion due to the cosmic expansion. When the quantum system is strongly 
bounded, it can be even not sensitive on the cosmic expansion.  It is e.g. the 
case for the Hydrogen atom which energy levels remain unaffected by the cosmic 
expansion. For the free condensate, the spatial part (in coordinate variables) decouples
from the temporal dependence.  Therefore if the wavefunction has some characteristic
scale, let say $r_{\text{x}}$, the the corresponding physical scale will be accordingly 
$R_{\text{x}}=a r_{\text{x}}$. Therefore whole physical scales of the wavefunction will 
expand homogeneously. In order to find the value of this growth we have to determine 
the total expansion of the universe since the formation of axion condensate till now ($a(t_0)=1$).
The axion condensate was formed at some $a(t_1)$ what we take to be $T = 1$ GeV. 
The corresponding value of time is $t_1 \simeq 2\cdot 10^{-7}$ sec. Based  on this, we have 
\begin{equation}
\frac{a(t_0)}{a(t_1)} = \frac{1 \text{GeV}}{T_{\text{rec}}} (1+z_{\text{rec}}) \simeq 5 \cdot 10^{12},
\end{equation}  
where the $T_{\text{rec}}=0.2$ eV is the energy scale of recombination and 
$z_{\text{rec}} \simeq 10^3$ is the corresponding value of redshift. So if the present
characteristics size of the axion wavefunction is of order of the galactic scale 
$R_{\text{x}} = 10$ kpc, it corresponds to $R_{\text{x}} \simeq 10^8$ m  at $t_1$.  

Let us examine firstly the case without the vortex, $l=0$. The solution of Gross-Pitaevski 
equation is of the form
\begin{equation}
\phi({\bf r},t)= \frac{C}{\sqrt{4\pi}}\frac{1}{a^{3/2}} \exp{\left\{-\frac{i}{\hslash} \int \frac{\mu}{a^2(t)} dt \right\}} \frac{\sin{(\sqrt{\lambda}r)}}{\sqrt{\lambda}r}.
\end{equation}
This solution describes a spherically symmetrical halo with the density
distribution 
\begin{equation}
\rho(r,t) = m N |\phi({\bf r},t)|^2 = \frac{m N |C|^2}{4\pi a^{3}} \left[ \frac{\sin{(\sqrt{\lambda}r)}}{\sqrt{\lambda}r} \right]^2.
\end{equation}
The density in the central region is given by 
\begin{equation}
\rho(r \rightarrow 0,t) = \frac{m N |C|^2}{4\pi a^{3}} := \rho_0 .  
\end{equation}
We must remember that the coordinate $r$ in not a physical distance, which is
$R=a \cdot r$. For a spherical
distribution we calculate the velocity rotation curve from the relation     
\begin{equation}
v(R)=\sqrt{\frac{G\mathcal{M}(R)}{R}},
\label{vel}
\end{equation}
where $\mathcal{M}(R)$ is the mass function and is expressed as
\begin{equation}
\mathcal{M}(R) = 4\pi  \int_0^{R} R'^{2} \rho(R') dR'.
\end{equation} 
Based on this, we find
\begin{equation}
v(R)=v_0 \sqrt{1-\frac{\sin(R/R_0)}{R/R_0}} \label{vR0}
\end{equation}
where
\begin{eqnarray}
v_0 &=&  \sqrt{2\pi G \rho_0 \frac{a^2}{\lambda}} =  \hslash \sqrt{\frac{\pi G \rho_0}{mE}},  \label{v00}\\
R_0 &=& \frac{a}{2\sqrt{\lambda}} = \frac{\hslash}{2\sqrt{2mE}}. \label{R00}
\end{eqnarray}

\begin{figure}[ht!]
$\begin{array}[c]{cc}
\includegraphics[width=6cm]{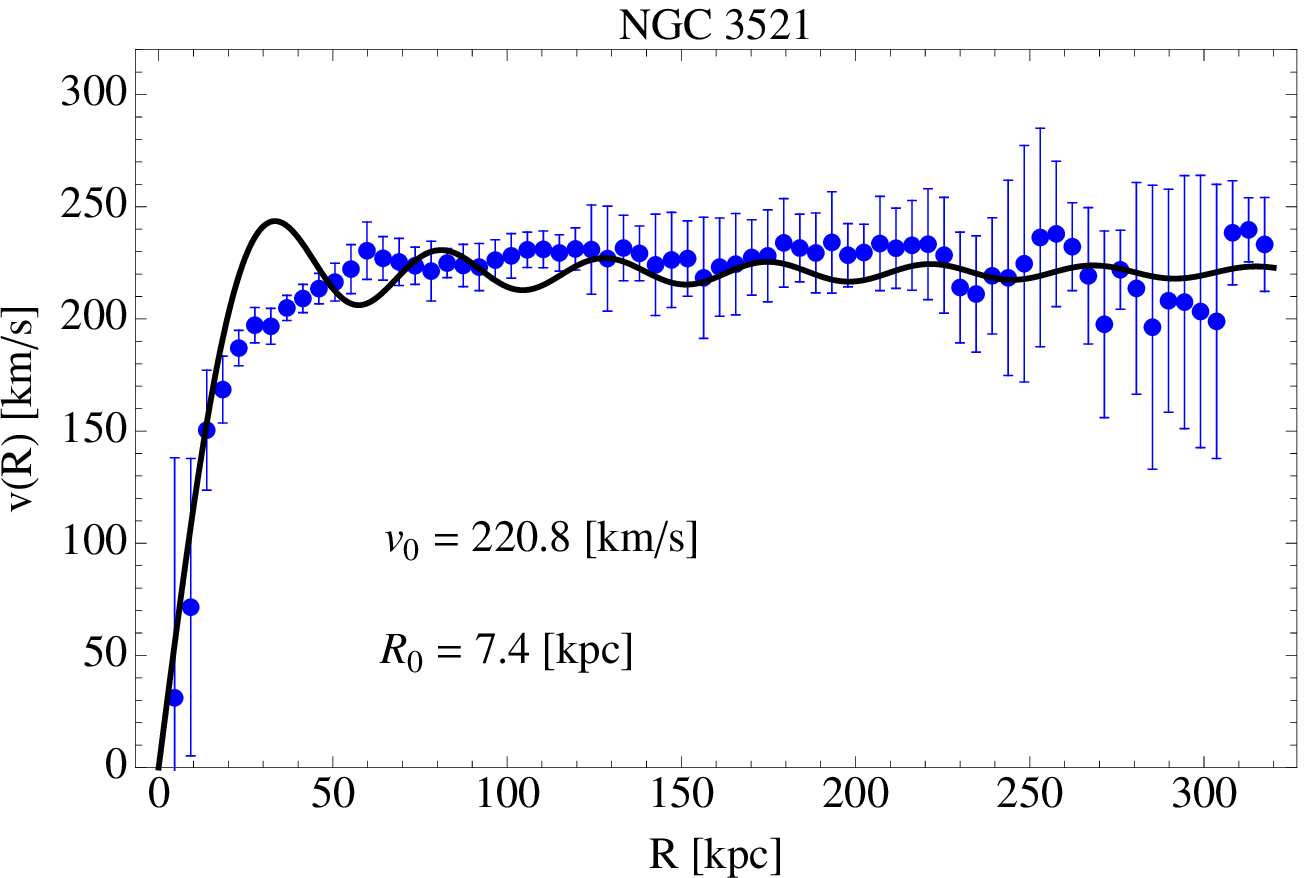} &  \includegraphics[width=6cm]{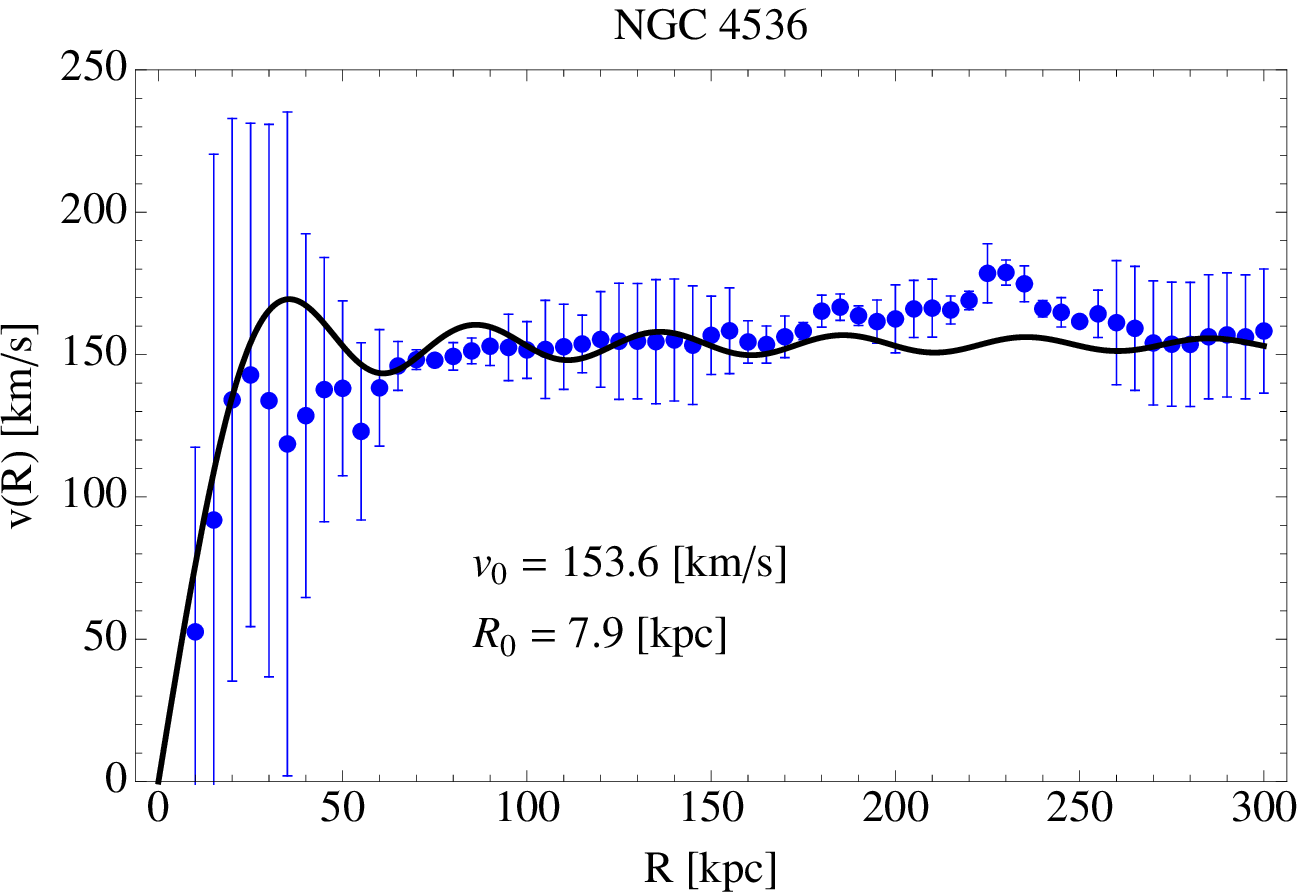}  \\
\includegraphics[width=6cm]{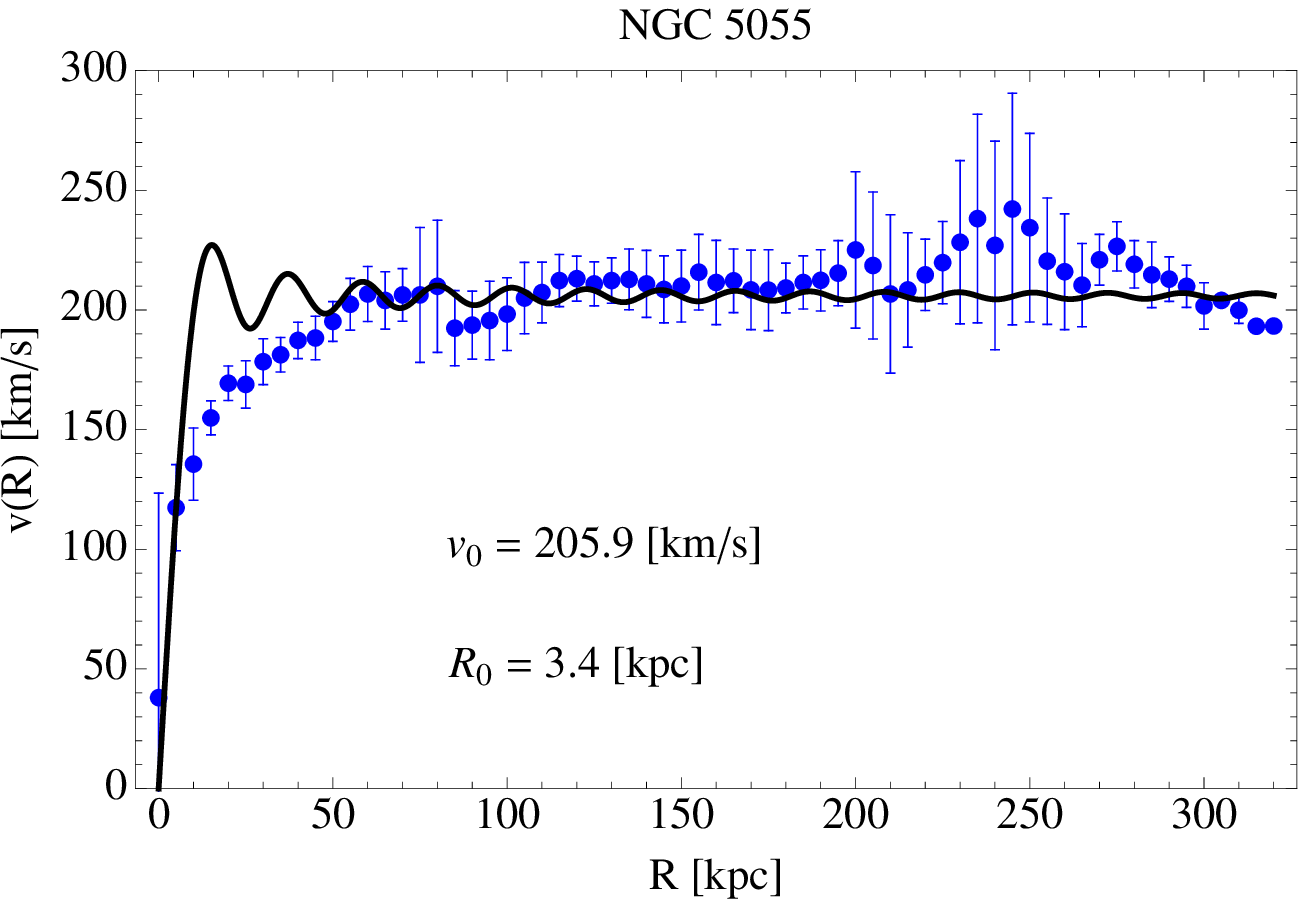}  & \includegraphics[width=6cm]{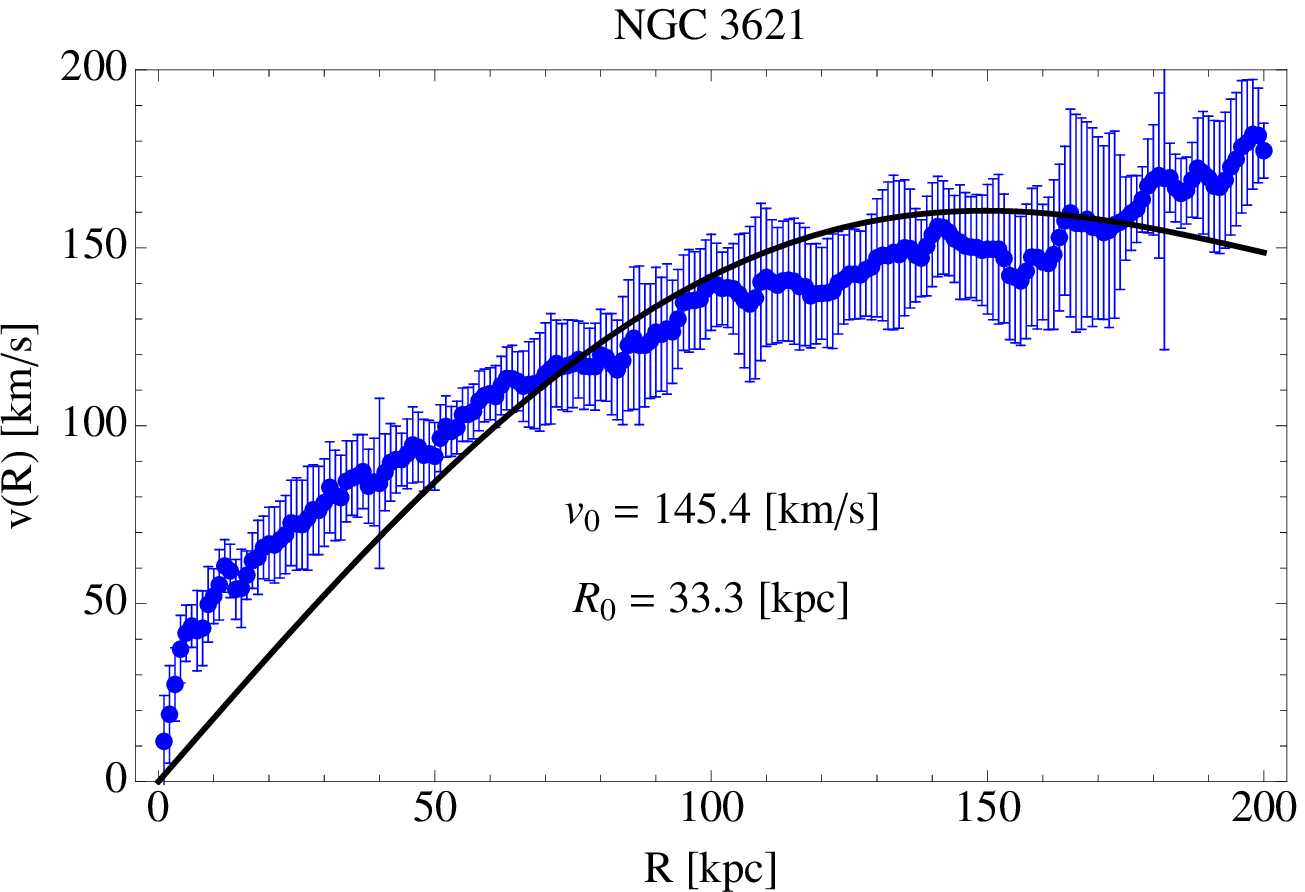}  
\end{array}$
\caption{The contribution to the galactic velocity curves given by the axion condensate. The 
data points are galactic velocity curves of the spiral galaxies. The NGC3521, 
NGC4536, NGC3621 are barred spiral galaxies (SBbc). The NGC5055 is the spiral galaxy with 
the central bulge (SbSc).}
\label{dist1}
\end{figure}

Figure~(\ref{dist1}) shows the velocity curve for axion condensate (\ref{vR0}).
We compare the theoretically predicted curves with the exemplary galactic 
rotation curves \cite{vortex:rvc}.   

This result describes the observed galaxy velocity curves quite well -- we can
see the characteristic change to a plateau in the rotation curves. This is not 
not seen for comparison with  NGC3621 galaxy, however even in this case,
the main observational features can be reproduced. 

Kinetic energy of the axion particles is equal to $E=\frac{m\sigma^2}{2}$, where 
$\sigma$ is the velocity dispersion. Based on this, the equations (\ref{v00}) and (\ref{R00}) 
can be rewritten to the following form   
\begin{eqnarray}
v_0 &=& \frac{\hslash}{m\sigma} \sqrt{2\pi G \rho_0}, \\
R_0 &=& \frac{1}{2} \frac{\hslash}{m\sigma}. \label{R0s}
\end{eqnarray}
Therefore, based on the observationally determined values of $R_0$ and $v_0$, the  
quantities $\rho_0$ and the product $m \sigma$ can be determined. However, the 
values of of $m$ and $\sigma$ cannot be determined independently. Combining the 
above equations, we find the consistency relation for our model 
\begin{equation}
\rho_0 = \frac{1}{8\pi G} \frac{v_0^2}{R_0}.
\end{equation}
Therefore, the energy density in the central region of the halo can be determined 
based on $R_0$ and $v_0$ determined form the observations. This quantity can 
be compared with the observational value determined independently.  

The value of velocity dispersion of axions can be expressed in the following way
\begin{equation}
\sigma \approx \frac{\hslash}{m c t_1}   \frac{a(t_1)}{a(t)}. \label{sigmaa}
\end{equation}
This expressions comes from the assumption that the axion momenta at the 
time $t_1$ is comparable with the corresponding Hobble factor.
Therefore $p=m \sigma \approx H = \frac{1}{2t_1}$, the factor $\frac{a(t_1)}{a(t)}$ 
is due to the cosmological redshift. The expression does not take into account 
the contribution due to gravitationally  induced thermalization. The interactions
between the axions are neglected here.  Based on this, the expression on the 
parameter $R_0$ takes the form 
\begin{equation}
R_0 \approx \frac{c t_1}{2} \frac{a(t)}{a(t_1)}.
\end{equation}
Based on this, the present value of $R_0$ is predicted to be  $R_0\approx 10^{-2}$pc. 
This value is six orders of magnitude shorter than the typically observed value 
$R_0\approx 10$ kpc. Therefore in order explain the galactic rotation curves by the
axion condensate, the velocity dispersion $\sigma$ (or the product $\sigma$ if the mass is 
not fixed) must be smaller than predicted from (\ref{sigmaa}). Namely for $R_0\approx 10$kpc,
based on (\ref{R0s}) with $m_a \approx 10^{-5}$eV, we find  $\sigma \approx 10^{-12} \frac{m}{s}$.
Therefore, the dispersion of the axions velocities must be extremely low in order 
to explain the galactic halo in terms of the wavefunction of the axion condensate.
This velocity is much smaller than this predicted from equation (\ref{sigmaa}). Moreover,
the additional velocity dispersion can be produced due the the gravitational interactions.
Therefore we conclude that, the axions of the mass order of  $m_a \approx 10^{-5}$eV, 
cannot form the quantum compact structures on the galactic scales. They can however 
form the structures on the smaller scales. 

Based on (\ref{R0s}) one can answer, what should be the mass of the axion-like particle
necessary to produce the  quantum condensate on the galactic scales. In order to answer 
this we have to assume some velocity dispersion of the dark matter particles. If they thermalize
due to the gravitational interactions, the velocity dispersion should be order of the radial velocity
of the stars (as in the isothermal sphere model). Taking $\sigma \approx 10^{2} \text{km}/\text{s}$ 
and $R_0\approx 10$kpc we find 
\begin{equation}
mc^2 = \frac{\hslash c}{2 R_0} \frac{c}{\sigma} \approx 10^{-30} \text{eV}. 
\end{equation} 
The particle must me therefore ultra-light. The interpretation of this result is clear 
when we recognize the right side in equation (\ref{R0s}). It is just, up to the numerical factor, 
the de Broglie wavelength $\lambda_{\text{dB}}$ of the axion-like particle. These wavelengths 
have to be of the galactic sizes in order to keep the coherence of the quantum system. Otherwise 
the whole structure will disintegrate due to decoherence. Here this condition naturally emerges.  
Namely, the equation (\ref{R0s}) can be rewritten into the form  
\begin{equation}
\lambda_{\text{dB}} = 4\pi R_0,
\end{equation}
where the right side is of the order of galactic diameter.  

\section{Vortex in the free axion condensate}

The similar results can be obtained considering the vortex solution with $l=1$. 
The velocity curves also exhibit a plateau, like the solution with $l=0$. The only 
difference is that we now have the angle dependence $\rho \sim cos^2 \theta$. 
This produces, like for each vertex solution, distortions from the spherical shape 
of a dark matter halo.

\begin{figure}[ht!]
\includegraphics[width=8cm]{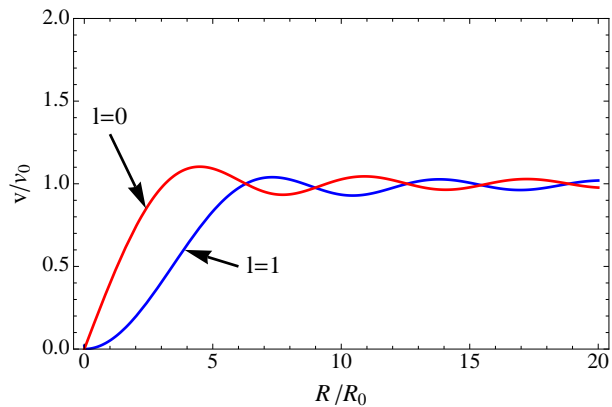} 
\caption{Comparison of the radial velocity curves for $l=1$ and $l=0$. In both case, the velocity curves are flat, 
$v(R)\approx v_0$, for $R\gg R_0$.}
\label{rvc}
\end{figure}

\begin{figure}[ht!]
\includegraphics[width=8cm]{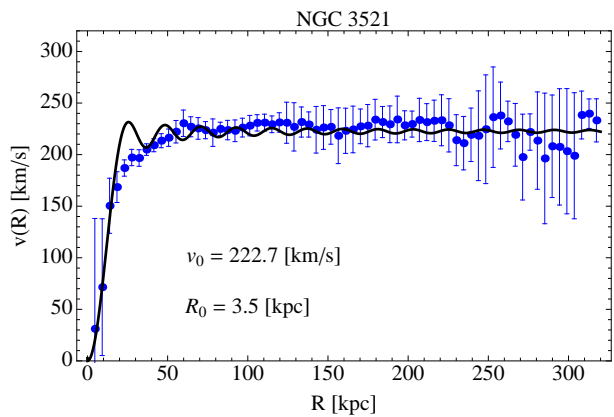} 
\caption{Radial velocity curve (\ref{vl1}) for the axion condensate with $l=1$ compared with the 
observational data for the galaxy NGC 3521. In general, we observe a good agreement between the 
theoretical curve and the observational data. The main difference is the first peak in the theoretical 
curve. The predicted radial velocity is at this point about $30\ \text{km}/\text{s}$ higher than the observed 
value.}
\label{3521v}
\end{figure}

The $l=1$ is a stable vortex configuration. Taking 
\begin{equation}
Y_1^1(\theta,\varphi)=\frac{1}{\sqrt{2\pi}}\frac{\sqrt{6}}{2}\cos\theta e^{i\varphi},
\end{equation}
we find the corresponding solution 
\begin{equation}
\phi({\bf r},t)= \frac{C\sqrt{3}}{2\sqrt{\pi}}\frac{1}{a^{3/2}} \exp{\left\{-\frac{i}{\hslash} \int \frac{\mu}{a^2(t)} dt \right\}} \left(
\frac{\sin{(\sqrt{\lambda}r)}}{(\sqrt{\lambda}r)^2}-\frac{\cos{(\sqrt{\lambda}r)}}{\sqrt{\lambda}r}
\right) \cos(\theta)e^{i\varphi}.
\end{equation}
In similarity with the $l=0$ case we find the density distribution 
\begin{equation}
\rho(r,\theta, t) = 3\rho_0  \underbrace{\left[ \frac{\sin{(\sqrt{\lambda}r)}}{(\sqrt{\lambda}r)^2}-
\frac{\cos{(\sqrt{\lambda}r)}}{\sqrt{\lambda}r}  \right]^2}_{=j^2_1\left(\sqrt{\lambda}r\right)} \cos^2 \theta.
\end{equation}
We have used here the definition of $\rho_0$ introduced in the $l=0$ case. However 
here one cannot interpret it as a density for $r\rightarrow 0$, because now  $\rho(r \rightarrow 0)=0$.
Note, that in the case considered, the density distribution strongly depends on the angle
$\theta$, namely $\rho \propto \cos^2\theta $. Therefore, at the surface of galactic disc ($\theta=\pi/2$) 
the density of dark matter halo vanishes. This dependence must be taken into account while 
calculating the corresponding velocity curves. In general, one should firstly calculate the 
surface density $\Sigma$, integration density  $\rho(r,\theta, t)$ over the $z$-direction.  Defining 
$\mathcal{R}$ as a physical radial variable on the surface $z=0$ we find
\begin{equation}
\Sigma(\mathcal{R}) = 3\rho_0 \int_{-\infty}^{+\infty}dz j^2_1\left(\frac{\sqrt{\lambda}}{a} 
\sqrt{z^2+\mathcal{R}^2}\right) \frac{z^2}{z^2+\mathcal{R}^2}. 
\end{equation}
We were however unable to compute this integral analytically. Based on this one can  
derive the mass function   
\begin{equation}
\mathcal{M}(\mathcal{R}) = 2\pi  \int_0^{\mathcal{R}} \mathcal{R}' \Sigma(\mathcal{R}') d\mathcal{R}'.
\end{equation} 
and then the corresponding velocity curve
\begin{equation}
v(\mathcal{R})=\sqrt{\frac{G\mathcal{M}(\mathcal{R})}{\mathcal{R}}}.
\end{equation}
Since this method is rather not practical computationally, we can alternatively 
integrate over the angular part, then 
\begin{equation}
\rho(r, t) = \int \rho(r,\theta, t) d\Omega = 4\pi \rho_0\left[ \frac{\sin{(\sqrt{\lambda}r)}}{(\sqrt{\lambda}r)^2}-
\frac{\cos{(\sqrt{\lambda}r)}}{\sqrt{\lambda}r} \right]^2.
\end{equation} 
With use of expression on $\mathcal(R)$ (without the factor $4\pi$, since we have already 
integrated over the solid angle) and  (\ref{vel}) we find 
\begin{equation}
v(R)=v_0 \frac{R_0}{R} \sqrt{4\left[ \cos(R/R_0)-1\right]+(R/R_0)^2+(R/R_0)\sin(R/R_0)}, \label{vl1}
\end{equation}
where $v_0$ and $R_0$ are defined in the same way as in the $l=0$ case. In Fig. \ref{rvc} we compare 
expression (\ref{vl1}) with the analogous function previously found in the $l=0$ case.
As we see there is no qualitative difference between the velocity curves for $l=1$ and $l=0$.
In both cases, the velocity curves become flat, $v(R)\approx v_0$, for $R\gg R_0$. 
In Fig. \ref{3521v} we compare velocity curve (\ref{vl1}) with the exemplary galactic velocity 
curve.
The fit gives a qualitatively good explanation of the observational data. Despite this, as we have discussed
before, the quantum structure on the galactic scales can be formed only if the axion-like particle is extremely low.
This results naturally  apply for the case of single vertex considered in this section. The vortex in axion condensate 
cannot forms on the galactic scales because the axion mass is to high. However, the vortices can be formed on the 
much smaller scales. As we have previously estimated, $R \sim 10^{-2}$pc, neglecting the gravitational 
interactions of the axion particles. Such a small vortices can however give seed to the galaxy formation. They can 
also be a source of the angular momenta of the primordial galaxies. This possibility  is an attractive subject for
the further studies. Therefore the concept of axion vortices can be still potentially useful in explaining formation of the 
galaxies.   

\section{Discussion}

We have considered axion condensate as a candidate for dark matter in galactic
halos. Condensates in laboratories are known to exhibit vortices induced by
rotation of the environment. Such vortices are unstable and decay into states
with the lowest angular momentum. We prolong this picture onto the entire 
Universe whose global rotation might give rise to the rotation of particular galaxies.
Precisely, the total angular momenta can decay onto the elementary vortices,
each with the same topological charge $l=1$.   

We have considered the non-normalizable states of the axion condensate with 
$l=0$ and $l=1$(vortex). It was shown that these solutions leads to the flat velocity 
curves characterized by oscillations in a plateau region. There was no qualitative 
difference between the velocity curves obtained in these two cases. We have 
confronted the theoretical curves with the  exemplary observational data. We found that
the theoretical curves reproduce the observational data. However, we have to keep 
in mind that we neglected contribution from the ordinary luminous matter. Therefore obtained 
velocity curves give only the contribution form the dark matter halo. The contribution 
from the luminous matter can be important near the galactic center, but is secondary 
at the larger distances. Therefore, the flatness of the velocity curves at $R\gg R_0$ will be 
conserved even in presence of this additional matter content.

In this paper we have restricted to the case of the free axion condensate. Some tiny interaction
can be interaction  present due to the gravitational forces. These interaction 
can be in fact important from the perspective of thermalization of the axion condensate. It 
requires further numerical studies to take into account the gravitational interactions in the 
considered scenario. In particular, the axion condensate will not expand freely during the 
cosmological evolutions, as it was observed in case of the free axion condensate.      
 
Another issue is that the axions have some velocity dispersion. Because of this, the individual 
wavefunctions of the axions differ a bit, and the wavefunction of the condensate should 
be considered as a superposition
\begin{equation}
\Psi({\bf x},t) = \int d^3{\bf k}  f(k) \phi_{k}({\bf x},t), 
\end{equation}  
where $\phi_k({\bf x},t)$ are eigenfunctions of the Hamiltonian. The function $f(k)$ is the Gauss-shaped 
function centered at $k=0$ and with the dispersion related to $\sigma$. This superposition 
of non-normalizable states will lead to the normalizable state $\Psi({\bf x},t)$.

The considered model has however serious limitations if applied to axions. Namely, the velocity
dispersion of the axions must be much smaller that predicted from the theory. Moreover, the gravitational
interaction of the axion will lead to the velocity dispersion much higher than required to form 
the quantum vortex on the galactic scales.  Intuitively, the de Broglie wavelength of axions is 
much smaller than the galactic scales. Therefore the quantum condensate will not keep the 
coherence on such scales. However, it does not exclude the axions as dark matter particles 
forming the galactic halo. Due to the gravitational interactions they thermalize and can be described
by the model of isothermal  sphere. In this case, the flat velocity curves are also explained, and 
the constant velocity of galaxies is related to the velocity dispersion of axions by the relation 
$v_{\text{c}}=\sqrt{2} \sigma$. The vortices in axion condensate can be however important in the 
process of the galaxy formation. Namely they can give the initial conditions for the galaxy formation.
In particular, they can be responsible for the initial angular moment.  This may also explain why the 
masses of the spiral galaxies are similar and equal to $\sim 10^{11} \ M_{\odot}$. 
This can be possibly explained since all the vortices have the same topological charge $l=1$, 
what is directly a quantum effect.  Therefore, initially, galaxies can grow on the independent 
vortices in the axion condensate. But later, when the accretion of matter (including other axions) 
proceed, the length of coherence of the individuals vortices became much smaller than the length 
of created gravitating structure. The vortices in the axion condensate can be therefore a seeds of the galaxies
but have no important influence on them at the further stages.  However, the idea that the 
vortices  in the axion condensate could have something to do with the galaxy formation must me 
approved be the further investigations. 

Alternatively, the coherent quantum structure as the vortices, can be formed on the galactic distances
but only if the mass of the axion-like particles is extremely low, namely  is of the order of  $10^{-30}$eV.
In such a case  the de Broglie wavelength of these particles is comparable with the galactic sizes. 
This condition prevents the disintegration of the quantum system due to the lack of coherence 
between the different parts of the quantum galactic halo. Concluding, the explanation 
of the galactic velocity curves presented in this paper can still holds, but only if the dark matter 
particles are ultra-light. The axions of this mass are however illicit. 
 
\section*{Acknowledgments}

This work was supported in part by the Marie Curie Actions Transfer of
Knowledge project COCOS (contract MTKD-CT-2004-517186).

\end{document}